\documentclass[twocolumn]{aastex63}

\received{XX XX, 20XX}
\revised{XX XX, 20XX}
\accepted{XX XX, 20XX}
\submitjournal{ApJ}
														
\shorttitle{AGN in a massive quiescent galaxy at $z=3.09$}
\shortauthors{Kubo et al.}
 
\begin{document}

\title{An AGN with an ionized gas outflow in a massive quiescent galaxy in a protocluster at $\bf z=3.09$}

\correspondingauthor{Mariko Kubo}
\email{kubo@cosmos.phys.sci.ehime-u.ac.jp}

\author[0000-0002-7598-5292]{Mariko Kubo}
\affiliation{Ehime University, 2-5 Bunkyo-cho, Matsuyama, Ehime 790-8577, Japan}
\author[0000-0003-1937-0573]{Hideki Umehata}
\affiliation{Institute of  Advanced Research, Graduate School of Science, Nagoya University, Furo-cho, Chikusa-ku, Nagoya, Aichi 464-8601, Japan}
\affiliation{Institute for Cosmic Ray Research, The University of Tokyo, 5-1-5 Kashiwa-no-Ha, Kashiwa City, Chiba, 277-8582, Japan}
\author[0000-0003-1747-2891]{Yuichi Matsuda}
\affiliation{National Astronomical Observatory of Japan, 2-21-1 Osawa, Mitaka, Tokyo 181-8588, Japan}
\affiliation{Department of Astronomy, Graduate School of Science, The University for Advanced Studies, SOKENDAI, Mitaka, Tokyo 181-8588, Japan}
\author[0000-0002-1732-6387]{Masaru Kajisawa}
\affiliation{Ehime University, 2-5 Bunkyo-cho, Matsuyama, Ehime 790-8577, Japan}
\author[0000-0002-4834-7260]{Charles C. Steidel}
\affiliation{Cahill Center for Astronomy and Astrophysics, California Institute of Technology, MC 249-17, Pasadena, CA 91125, USA}
\author{Toru Yamada}
\affiliation{Institute of Space and Aeronautical Science, Japanese Aerospace Exploration Agency, 3-1-1, Yoshinodai, Chuo-ku, Sagamihara, Kanagawa, 252-5210, Japan}
\author[0000-0002-4937-4738]{Ichi Tanaka}
\affiliation{Subaru Telescope, National Astronomical Observatory of Japan, 650 North A’ohoku Place, Hilo, HI 96720, U.S.A.}
\author[0000-0001-6469-8725]{Bunyo Hatsukade}
\affiliation{Institute of Astronomy,  Graduate School of Science, The University of Tokyo, 2-21-1 Osawa, Mitaka, Tokyo 181-0015, Japan}
\author[0000-0003-4807-8117]{Yoichi Tamura}
\affiliation{Department of Physics, Nagoya University, Furo-cho, Chikusa-ku, Nagoya, Aichi 464-8601, Japan}
\author[0000-0002-6939-0372]{Kouichiro Nakanishi}
\affiliation{National Astronomical Observatory of Japan, 2-21-1 Osawa, Mitaka, Tokyo 181-8588, Japan}
\affiliation{Department of Astronomy,  Graduate School of Science, The University for Advanced Studies, SOKENDAI, Mitaka, Tokyo 181-8588, Japan}
\author[0000-0002-4052-2394]{Kotaro Kohno}
\affiliation{Institute of Astronomy, Graduate School of Science, The University of Tokyo, 2-21-1 Osawa, Mitaka, Tokyo 181-0015, Japan}
\affiliation{Research Center for the Early Universe, The University of Tokyo, 7-3-1 Hongo, Bunkyo-ku, Tokyo 113-0033,Japan}
\author[0000-0003-4814-0101]{Kianhong Lee}
\affiliation{Institute of Astronomy, Graduate School of Science, The University of Tokyo, 2-21-1 Osawa, Mitaka, Tokyo 181-0015, Japan}
\author{Keiichi Matsuda}
\affiliation{Department of Physics, Nagoya University, Furo-cho, Chikusa-ku, Nagoya, Aichi 464-8601, Japan}
\author[0000-0003-3139-2724]{Yiping Ao}
\affiliation{Purple Mountain Observatory and Key Laboratory for Radio Astronomy, Chinese Academy of Sciences, Nanjing, China}
\affiliation{School of Astronomy and Space Science, University of Science and Technology of China, Hefei, China}
\author[0000-0002-7402-5441]{Tohru Nagao}
\affiliation{Ehime University, 2-5 Bunkyo-cho, Matsuyama, Ehime 790-8577, Japan}
\author[0000-0001-7095-7543]{Min S. Yun}
\affiliation{Department of Astronomy, University of Massachusetts, Amherst, MA 01003, USA}

\nocollaboration{16}


\begin{abstract}

We report the detection of an ionized gas outflow from an $X$-ray  
active galactic nucleus (AGN) hosted in a massive quiescent galaxy in a protocluster at $z=3.09$ (J221737.29+001823.4).
It is a type-2 QSO with broad ($W_{80}>1000$ km s$^{-1}$) and strong 
($\log (L_{\rm [OIII]}$ / erg s$^{-1})\approx43.4$) [O {\footnotesize III}]$\lambda\lambda$4959,5007 emission lines
detected by slit spectroscopy in three-position angles using 
Multi-Object Infra-Red Camera and Spectrograph (MOIRCS) on the Subaru telescope 
and the Multi-Object Spectrometer For Infra-Red Exploration (MOSFIRE) on the Keck-I telescope. 
In the all slit directions, [O {\footnotesize III}] emission is extended to $\sim15$ physical kpc
and indicates a powerful outflow spreading over the host galaxy.
The inferred ionized gas mass outflow rate is $\rm 22\pm3~M_{\odot}~yr^{-1}$. 
Although it is a radio source, according to the line diagnostics using H$\beta$, [O {\footnotesize II}], and [O {\footnotesize III}], 
photoionization by the central QSO is likely the dominant ionization mechanism rather than shocks caused by radio jets. 
On the other hand, the spectral energy distribution of the host galaxy is well characterized
as a quiescent galaxy that has shut down star formation by several hundred Myr ago.
Our results suggest a scenario that QSOs are powered after the shut-down of the star formation
and help to complete the quenching of massive quiescent galaxies at high redshift.
\end{abstract}



\section{Introduction}

Giant elliptical galaxies are the dominant population in clusters of galaxies today
and their formation history is one of the most important issues of observational cosmology. 
According to their tight color-magnitude relation, 
they are thought to have formed the bulk of their stars in the early Universe
(e.g., \citealt{1992MNRAS.254..601B,1998A&A...334...99K,1998MNRAS.299.1193B}). 
Massive galaxies with quiescent star formation similar to giant ellipticals today
have been now discovered at up to $z=4$
(e.g., \citealt{2018A&A...618A..85S,2019ApJ...885L..34T,2018ApJ...867....1K, 2021ApJ...919....6K,2020ApJ...889...93V,2020ApJ...903...47F, 2020ApJ...905...40S}).
They are characterized by significant Balmer/4000\AA~breaks and suppressed emission in the rest-frame UV and far-infrared (FIR). 
According to the detailed analysis of their star formation histories (SFH)
using multi-wavelength photometry and deep near-infrared (NIR) spectroscopic observations,
they have likely formed via a burst of star formation quenched suddenly, and evolved passively for several hundred Myr to a few Gyr ago; 
however, it is not yet understood how they were quenched and how they were maintained quiescence
when the Universe was still rich in gas. 

Several quenching mechanisms to suppress the star formation by removing or heating gas
have been adopted in cosmological numerical simulations to reproduce
massive quiescent galaxies in the early Universe (e.g., \citealt{2017MNRAS.465.3291W,2021MNRAS.506.4760D}). 
Active galactic nuclei (AGN) feedback is a plausible quenching mechanism 
(e.g., \citealt{2003ApJ...597..893N,2005A&A...437...69B,2006ApJS..163....1H,2016MNRAS.462.2418S}) 
leading to a well-established relationship between the masses of the supermassive black holes (SMBH) 
and the bulge luminosities/masses or velocity dispersions of their host galaxies
(e.g., \citealt{1998AJ....115.2285M,2000ApJ...539L...9F,2013ARA&A..51..511K}). 
AGN feedback is observed as quasar (radiative) mode that occurs around the peak of the AGN activity
and radio (kinetic or jet) mode that occurs at a low accretion rate \citep{2012ARA&A..50..455F}; 
strong outflows from AGNs have been observed 
(e.g., \citealt{2011ApJ...729L..27R,
2014MNRAS.441.3306H,2015A&A...580A.102C,2016MNRAS.456.1195H,2017ApJ...837...91B}) 
while radio-loud AGNs are often hosted by giant ellipticals in the local Universe
(e.g., \citealt{2005MNRAS.362...25B}).
But it is still unclear whether AGNs have quenched the star formation;
the star formation rates (SFR) of host galaxies of AGNs range widely 
(e.g., \citealt{2012A&A...540A.109S,2012A&A...545A..45R,2015MNRAS.452.1841S,2015MNRAS.453L..83M}).

Here we present a spatially extended ionized gas outflow from a type-2 AGN in a protocluster at $z=3.09$ 
detected with [O {\footnotesize III}]$\lambda\lambda$4959,5007 emission lines.
Its host galaxy is well characterized as a massive quiescent galaxy by detecting significant Balmer/4000 \AA~breaks photometrically. 
Thus the AGN in this particular object is an excellent target to understand how an AGN worked in a giant elliptical when it has been quenched star formation. 
This paper is organized as follows: 
In Section 2, we describe the target and observations. 
In Section 3, we describe the spectral energy distribution (SED) 
fitting and spectral analysis method. 
In Section 4, we present the result of the SED fitting and the spectral analysis. 
In Section 5, we discuss the properties of this QSO found in a quiescent galaxy and its powering mechanism. 
Section 6 summarizes our conclusions. 
In this study, we adopt cosmological parameters $\Omega_{\rm m} =0.3$, 
$\Omega_{\rm \Lambda} =0.7$ and $H_0 =70$ km s$^{-1}$ Mpc$^{-1}$.  
We assume a \citet{2003PASP..115..763C} Initial Mass Function (IMF). 
Magnitudes are expressed in the AB system. 
We use the vacuum rest-frame wavelengths of emission lines.

\begin{figure}
\gridline{\fig{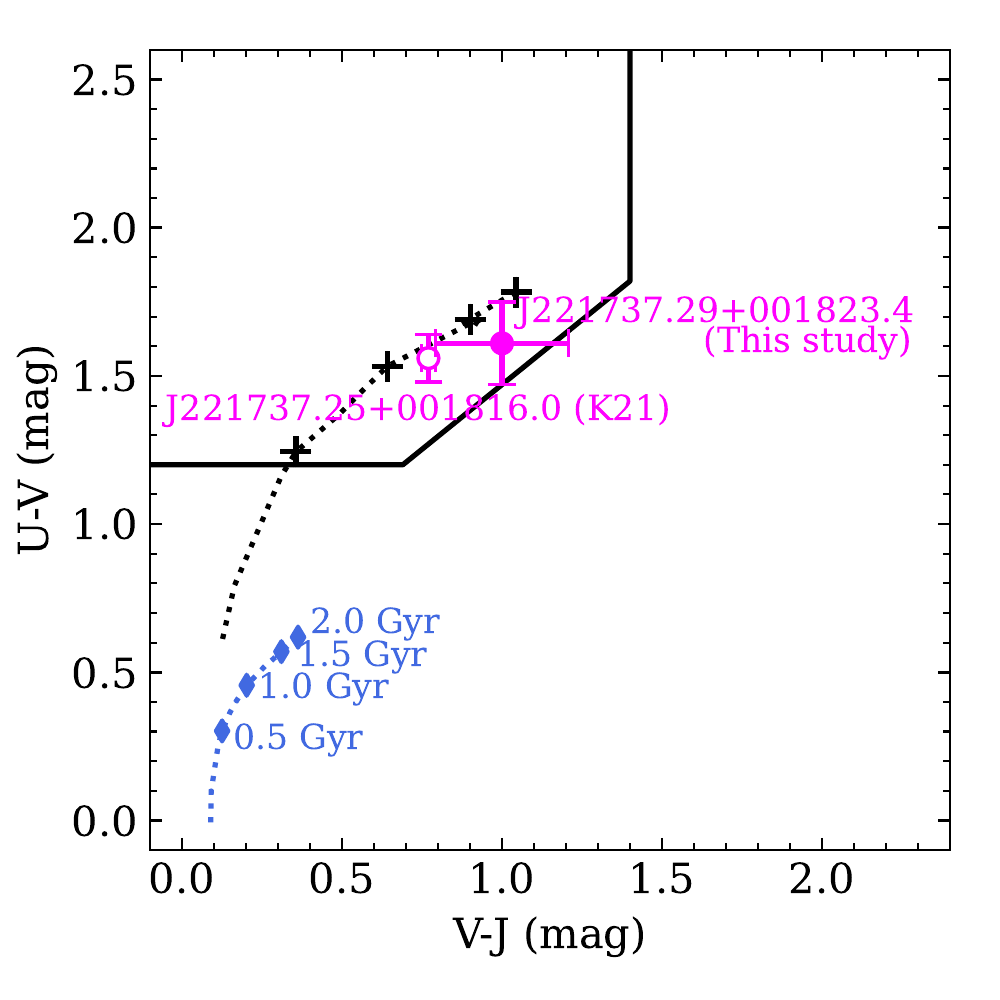}{0.4\textwidth}{}}
\caption{The rest-frame $UVJ$ color diagram. 
The magenta filled circle shows the target of this study (J221737.29+001823.4) 
and open circle shows a massive quiescent galaxy in the same protocluster
confirmed in \hyperlink{2021ApJ...919....6K}{K21} (J221737.25+001816.0).
The points and curves show the color evolution tracks for SED models 
with age between 0.1 to 2  Gyr computed with {\sf GALAXEV} \citep{2003MNRAS.344.1000B}. 
The black crosses with a dotted curve show the color evolution track 
for a single burst star formation model with $A_V=0$. 
The blue diamonds and curve show the color evolution track 
for a constant star formation model with $A_V=0$. 
The points are shown at ages 0.5, 1.0, 1.5, and 2.0 Gyr. 
The black solid line show the color criterion for quiescent galaxies at $2.0<z<3.5$ in \citet{2013ApJ...770L..39W}. 
\label{fig:uvj}}
\end{figure}

\begin{deluxetable}{lr}
\tablenum{1}
\tablecaption{Summary of the data \label{tab:data}}
\tablewidth{0pt}
\tablehead{ \colhead{Items} & \colhead{References}}
\startdata
Photometry & \\
\hline
CFHT MegaCam $u^{\star}$ & not published (PI: Cowie)  \\
Subaru S-Cam $BVRi'z'$ & \citet{2004AJ....128..569M} \\
                                                  &  \citet{2004AJ....128.2073H} \\
Subaru MOIRCS $JHK_s$ & \citet{2013ApJ...778..170K} \\
Subaru nuMOIRCS $K_s$ & \hyperlink{2021ApJ...919....6K}{K21}\\
{\it HST} ACS $F814W$ & in archive (PID9760) \\
$Spitzer$ IRAC \& MIPS 24$\mu$m & \citet{2009ApJ...692.1561W} \\
ALMA 1.2 mm & \citet{2018PASJ...70...65U} \\
ALMA 3.0 mm & \citet{2019Sci...366...97U} \\
VLA 6 GHz & Umehata et al. in prep \\
VLA 3 GHz & \citet{2017ApJ...850..178A} \\
VLA 1.4 GHz &  \citet{2004ApJ...606...85C} \\
{\it Chandra} 0.5-8 keV & \citet{2009MNRAS.400..299L} \\
\hline
Spectroscopy & \\
\hline
Subaru MOIRCS VPH-$K$ & \hyperlink{2015ApJ...799...38K}{K15} \\
Keck MOSFIRE $H$ and $K$ & \citet{2019Sci...366...97U} \\
VLT MUSE & \citet{2019Sci...366...97U} \\
\enddata
\end{deluxetable}

\begin{deluxetable}{lr}
\tablenum{2}
\tablecaption{Photometry \label{tab:phot}}
\tablewidth{0pt}
\tablehead{ \colhead{Items} & \colhead{Values}}
\startdata
$u^{\star}$  (mag)    & $>27.2~(2\sigma)$\\
$B$  (mag)              & $27.28\pm0.26$ \\
$V$  (mag)              & $26.77\pm0.20$ \\
$R$   (mag)              & $26.57\pm0.18$ \\
$i'$  (mag)                & $26.12\pm0.18$ \\
$z'$  (mag)                & $25.79\pm0.21$ \\
$F814W$ (mag)           & $26.66\pm0.32$ \\
$J$  (mag)            & $>24.58~(2\sigma)$ \\
$H$  (mag)          & $>24.66~(2\sigma)$ \\
$K_s$  (mag)       & $22.58\pm0.08$ \\
3.6 $\mu$m (mag)    & $22.15\pm0.06$ \\
4.5 $\mu$m (mag)    & $21.92\pm0.05$ \\
5.8 $\mu$m (mag)    & $21.86\pm0.16$ \\
8.0 $\mu$m (mag)    & $21.67\pm0.18$ \\
\hline
24 $\mu$m ($\mu$Jy) & $<100$ (3$\sigma$) \\
1.2 mm ($\mu$Jy) & $<75~(3\sigma)$\\
3.0 mm ($\mu$Jy) & $<43~(3\sigma)$\\
6 GHz ($\mu$Jy) & $9.56\pm0.85$\\
3 GHz ($\mu$Jy) & $13.07\pm1.64$ \\
1.4 GHz ($\mu$Jy) &  $<48~(4\sigma)$\\
0.5-2 keV (erg cm$^{-2}$ s$^{-1}$) & $1.0^{+0.6}_{-0.4}\times10^{-16}$\\
2-8 keV (erg cm$^{-2}$ s$^{-1}$) & $11.2^{+4.3}_{-3.3}\times10^{-16}$\\
\enddata
\end{deluxetable}

\section{Target and data} \label{sec:targetdata}
\subsection{Target}\label{subsec:target}

Table \ref{tab:data} summarizes the data and their reference of the target.
Our target is one of the galaxies confirmed by our NIR spectroscopic observations
of massive galaxies in a protocluster at $z=3.09$ in the SSA22 field using multi-object infrared camera and spectrograph 
(MOIRCS; \citealt{2006SPIE.6269E..16I,2008PASJ...60.1347S}) 
on the Subaru telescope (\citealt{2015ApJ...799...38K} hereafter \hyperlink{2015ApJ...799...38K}{K15}; referred J221737.3+001823.2). 
Hereafter we call the target as J221737.29+001823.4 adopting the source position 
measured on the new $K_s$-band image taken with updated MOIRCS (nuMOIRCS)\citep{2021ApJ...919....6K} (hereafter \hyperlink{2021ApJ...919....6K}{K21})
using {\sf SExtractor} \citep{1996A&AS..117..393B}. 
The source is an AGN matching with the $X$-ray source at 0.18 arcsec angular separation 
in the catalog obtained with {\it Chandra} in \citet{2009ApJ...691..687L,2009MNRAS.400..299L}. 
Its redshift is confirmed by detecting the [O {\footnotesize III}]$\lambda\lambda$4959,5007 emission   
which shows two peaks corresponding to $z=3.0851 \pm 0.0001$ and $3.0926 \pm 0.0003$ 
indicating complex kinematics of the ionized gas. 

As we noted in \hyperlink{2015ApJ...799...38K}{K15} and revisited in section \ref{subsec:sed}, 
the SED of J221737.29+001823.4 is well-fitted by a quiescent galaxy. 
Figure \ref{fig:uvj} shows the rest-frame $UVJ$ color diagram to select quiescent galaxies (e.g., \citealt{2013ApJ...770L..39W}). 
The $UVJ$ colors are interpolated from the best-fit SED described in section 3.1. 
J221737.29+001823.4 is classified as a quiescent galaxy similar to 
another massive quiescent galaxy with $z_{\rm spec}=3.0922^{+0.008}_{-0.004}$ 
at R.A., Dec = 22:17:37.25, +00:18:16.0, only 7.5 arcsec ($\approx60$ in physical kpc) 
away from the target confirmed in \hyperlink{2021ApJ...919....6K}{K21}. 
We note that J221737.25+001816.0 is not detected in [O {\footnotesize III}] emission and the {\it Chandra} and radio data described below, 
but shows the weak [O {\footnotesize II}] emission that can originate in an AGN.
Including our studies, several studies have shown the prevalence of massive quiescent galaxies
in protoclusters at up to $z=3.37$
(e.g., \citealt{2013ApJ...778..170K}; \hyperlink{2021ApJ...919....6K}{K21}; \citealt{2021ApJ...911...46S,2021arXiv210907696M}).
Such massive quiescent galaxies in protoclusters are the most plausible progenitors of typical giant ellipticals today. 

In the SSA22 protocluster, AGNs have been surveyed using the {\it Chandra}
and also {\it Spitzer} data \citep{2009ApJ...692.1561W},
and those hosted by galaxies ranging from Ly$\alpha$ emitters to sub-mm galaxies have been studied
(\citealt{2009ApJ...692.1561W,2009ApJ...691..687L,2009MNRAS.400..299L,2009ApJ...700....1G,2010ApJ...724.1270T,2013ApJ...778..170K}; 
 \hyperlink{2015ApJ...799...38K}{K15}; \citealt{2015ApJ...815L...8U,2019Sci...366...97U,2021ApJ...919...51M}).
Among them, J221737.29+001823.4 is not particularly luminous in $X$-ray
but is the most [O {\footnotesize III}] luminous source observed at this point (\hyperlink{2015ApJ...799...38K}{K15}).

\subsection{Photometry}\label{subsec:data}

Table \ref{tab:phot} summarizes the measured magnitudes and fluxes of the target.
The $u^{\star}$ to 8.0 $\mu$m photometry was taken in the same way as \hyperlink{2021ApJ...919....6K}{K21}. 
The $u^{\star}$ to $K_s$, and $F814W$-band images are convolved to match the PSF to a FWHM
of $\approx1.0$ arcsec, and measured fluxes with a 2.0 arcsec diameter aperture.
The IRAC $3.6-8.0~\mu$m photometry is applied aperture correction
computed by using the $K_s$-band image to match with $u^{\star}$ to $K_s$-band (see detail in \citealt{2013ApJ...778..170K}).
Then we corrected the PSF matched photometries multiplying by total (Kron flux measured on nuMOIRCS image) to aperture photometry ratio in $K_s$-band.
Major emission lines shifting to the bandpasses were checked spectroscopically 
and subtracted from these broadband measurements.
[O {\footnotesize II}], H$\beta$ and [O {\footnotesize III}]
were subtracted from the $H$ and $K$-band fluxes, respectively, while H$\alpha$ and [N {\footnotesize II}] are shifted out of the bandpasses. 
Ly$\alpha$ and C{\footnotesize IV} were ignorable since they were observed
with the Multi-Unit Spectroscopic Explorer (MUSE) 
on the Very Large Telescope (VLT) (\citealt{2019Sci...366...97U}; 2 $\sigma$ 
$= 0.3\times10^{-18}$ erg s$^{-1}$ cm$^{-2}$ arcsec$^{-2}$) but are not detected.
Type-2 AGNs show strong Ly$\alpha$ in general (e.g., \citealt{2002ApJ...571..218N,2019A&A...626A...9M}).
Indeed, Ly$\alpha$ fluxes of a few $10^{-16}$ erg s cm$^{-2}$ s$^{-1}$ were detected in 
the narrow line AGNs at $z\sim3$ with [O {\footnotesize III}] fluxes similar to J221737.29+001823.4 \citep{2018ApJ...866..119L}.
Although J221737.29+001823.4 itself showed no strong Ly$\alpha$, 
it is likely associated with the extended Ly$\alpha$ nebulae which indicate the presence of abundant intergalactic gas \citep{2019Sci...366...97U}. 
Thus it is natural to consider that Ly$\alpha$ photons of J221737.29+001823.4 are
absorbed and/or scattered by circum/intergalactic media to form the extended Ly$\alpha$ nebulae (e.g., \citealt{2009ApJ...693L..49H})
though the detail of Ly$\alpha$ damping mechanism is beyond the scope of this study. 

J221737.29+001823.4 was observed by ALMA in the Band-6 and Band-3, which gives upper limits at 1.2 mm and 3 mm
\citep{2015ApJ...815L...8U,2017ApJ...835...98U,2018PASJ...70...65U,2019Sci...366...97U}.
J221737.29+001823.4 was also observed by Karl G. Jansky Very Large Array (VLA) C-band (6 GHz), 
S-band (3 GHz; \citealt{2017ApJ...850..178A}) and L-band (1.4 GHz; \citealt{2004ApJ...606...85C}), and is detected in 3 and 6 GHz. 
Here we briefly explain the recent C-band observations (details will be presented in Umehata et al. in prep). 
The field of view covers an entire field of ALMA deep survey field in the SSA22 \citep{2018PASJ...70...65U},
which includes J221737.29+001823.4.
Observations were carried out in the A and B configurations covering 4.2-8.2 GHz with a total on-source time of 89 hours. 
After the data reduction and imaging with VLA Common Astronomy Software Applications (CASA), 
the resultant map has a synthesized beam size of $0''.89 \times 0''.79$ (PA=24.71 deg) 
and a rms level 0.35 $\mu$Jy/beam at the phase center. 
We measure the 3 GHz and 6 GHz fluxes using CASA/imfit. 
We also put a point-source upper limit at 1.4 GHz.
We put an upper limit of {\it Spitzer} MIPS 24 $\mu$m \citep{2009ApJ...692.1561W} (see also \citealt{2013ApJ...778..170K}).
We use the {\it Chandra} $X$-ray flux values listed in \citet{2009MNRAS.400..299L}.
Its radio flux is around the detection limit of VLA COSMOS 3 GHz survey \citep{2017A&A...602A...1S}
and $X$-ray flux is lower than the detection limit of the {\it Chandra} COSMOS Legacy survey \citep{2016ApJ...817...34M}. 
To summarize, J221737.29+001823.4 is an object hardly detected by the deep and wide surveys to date.  

\begin{figure*}
\gridline{\fig{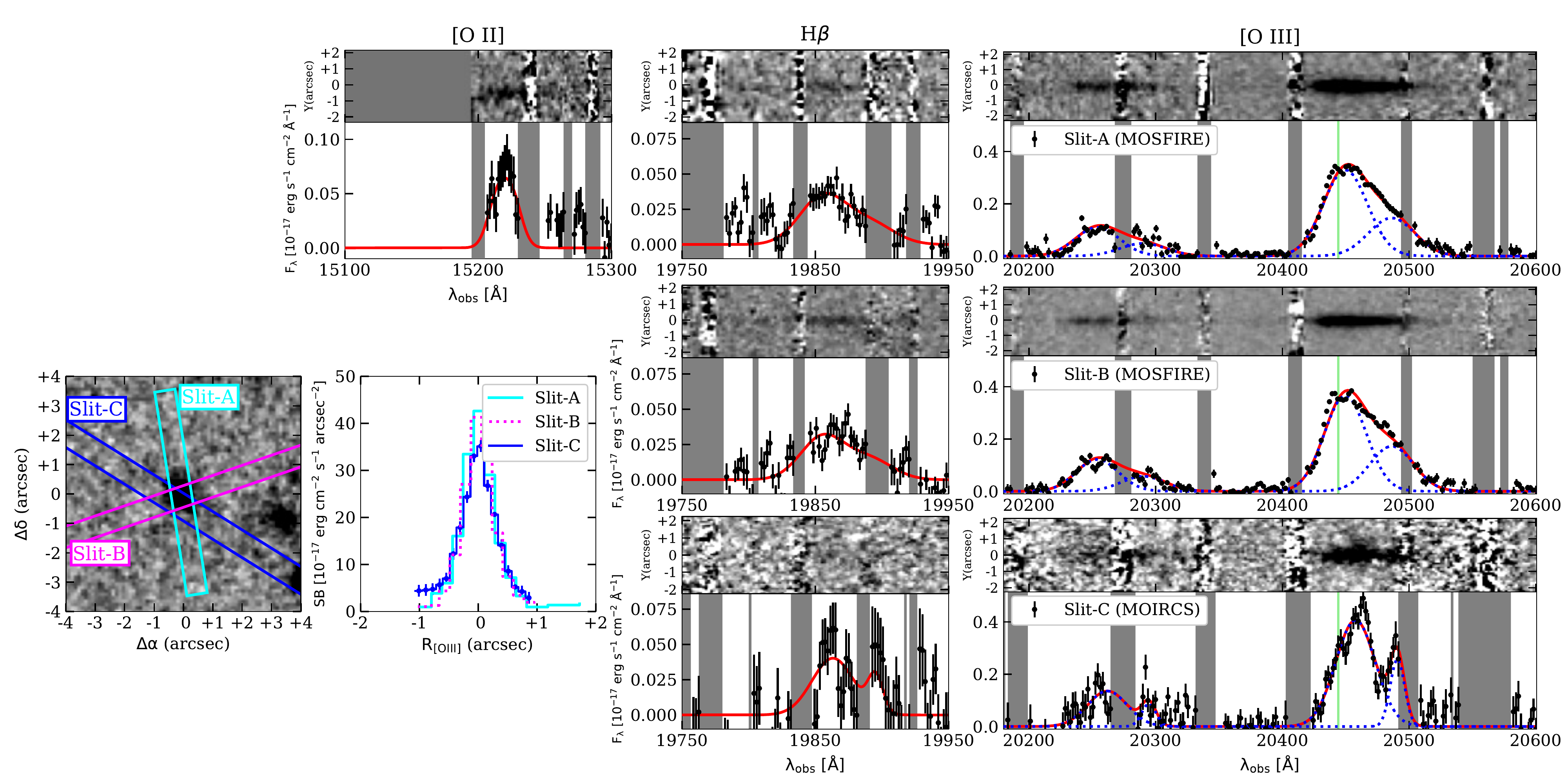}{0.99\textwidth}{}}
\caption{
Bottom left: The left panel shows $K_s$-band image of the target.
The image size is eight arcsec square. 
The cyan, magenta, and blue rectangles show the locations for slit-A, slit-B, and slit-C, respectively. 
The right panel shows the radial surface brightness profile of the [O {\footnotesize III}] emission line at each slit position. 
The north sides (upper side of left panel) of the slits are positive and the areas detected above 2$\sigma$ are plotted. 
Right: The spectral images and one-dimensional spectra. 
The top row shows the [O {\footnotesize II}], H$\beta$ and [O {\footnotesize III}] spectra observed at slit-A. 
The black points are the observed spectra in which the residual sky is subtracted. 
The red curves show the best-fit spectra. 
The gray shaded regions are masked in the spectral fittings.
The vertical green line shows the $z_{50}$ of [O {\footnotesize II}] corresponding to [O {\footnotesize III}].
The middle and bottom rows from central to right are similar to the top row
but show the H$\beta$ and [O {\footnotesize III}] spectra observed at slit-B and slit-C, respectively. 
\label{fig:spectra}}
\end{figure*}

\subsection{Spectrocopy}\label{subsec:data}

The NIR spectroscopic observations were performed
using MOIRCS (\hyperlink{2015ApJ...799...38K}{K15})
and Multi-Object Spectrometer For Infra-Red Exploration
(MOSFIRE; \citealt{2012SPIE.8446E..0JM}) on Keck-I telescope 
(\citealt{2019Sci...366...97U} and Umehata et al. in prep for details). 
We show the slit positions in Figure \ref{fig:spectra}. 
The observations were conducted using three slit position angles, 5 (Slit-A), 105 (slit-B), and 58 (slit-C) degrees. 
For slit-A and B, we conducted a $K$-band spectroscopy for 3.2 
and 3.3 h net exposure with MOSFIRE on 11 and 12th
September 2020 (\citealt{2019Sci...366...97U} and Umehata et al. in prep). 
For slit-A, an $H$-band spectrum was obtained with 2.5 h exposure 
with MOSFIRE on 11th September 2020 (see detail in \hyperlink{2021ApJ...919....6K}{K21}). 
The slit widths were $0.7$ arcsec for all the MOSFIRE observations. 
For slit-C, we conducted a $K$-band spectroscopy with 3.8 h exposure
with MOIRCS on Subaru telescope on 29th October 2012 (\hyperlink{2015ApJ...799...38K}{K15}). 
The location of slit-C was offset from the center of the target in order 
to observe the other two targets simultaneously. 
The slit width for the MOIRCS observation was $0.8$ arcsec.
All observations were conducted in good seeing conditions with FWHM PSF sizes $0.5$ to $0.8$ arcsec. 
Both MOIRCS and MOSFIRE observations were performed by 2-position mask-nod sequence dither along slits
to perform sky subtraction accurately. 
The calibration of the fluxes was performed using one Telluric standard star
at similar air masses taken before or after the observations each night.
The spectral resolutions for MOSFIRE $K$ ($H$) and MOIRCS $K$-band 
at the given configurations were $3620~(3660)$ and $1700$, respectively. 

Figure \ref{fig:spectra} shows the resulting spectra extracted by the MOSFIRE data reduction pipeline 
which is corrected of slit losses (multiplied by 1.53 to 1.98). 
[O {\footnotesize II}]$\lambda\lambda 3727,3729$, H$\beta$, 
and [O {\footnotesize III}] are detected significantly. 
Although the observing methods and conditions are not uniform, 
the H$\beta$ and [O {\footnotesize III}] fluxes measured at each slit location do not differ greatly. 
In the case of  J221737.25+001816.0 in \hyperlink{2021ApJ...919....6K}{K21} which was taken with the same masks as both slit-A and B, 
the differences between the slit loss corrected fluxes from MOSFIRE spectroscopy 
and the Kron fluxes based on the MOIRCS imaging in $H$ and $K_s$-band were $\sim20$ \%.
Thus the flux calibration in this study is almost correctly performed. 

\section{Analysis}\label{sec:analysis}
\subsection{SED fitting}\label{subsec:sed}

We used X-CIGALE \citep{2020MNRAS.491..740Y}, which models the SED 
of a host galaxy and $X$-ray to radio emission from an AGN simultaneously. 
The model parameters used in X-CIGALE are summarized in Table A\ref{tab:cigale}. 
Briefly, we included a stellar population synthesis model with
dust attenuation, dust emission, an optical to infrared emission model for an AGN,
X-ray power-law emission, and radio synchrotron emission.
For stellar population models, we adopted a
\citet{2003PASP..115..763C} IMF and a \citet{2003MNRAS.344.1000B} model with solar metallicity,
and dust attenuation law from \citet{2000ApJ...533..682C}. 
We assumed a delayed exponentially declining SFH model described as, 
$${\rm SFR}(t) \propto \frac{(t_o-t)}{\tau^2} \times \exp (-(t_o-t)/\tau)~{\rm for}~0~\leq~t~\leq~t_o$$
where $t$ is the lookback-time, $t_o$ is the lookback-time onset of star formation, 
and $\tau$ is the e-folding time at which the SFR peaks. 
We use the dust emission templates from \citet{2014ApJ...784...83D}, 
which are parameterized by a power-law slope $\alpha$ of $dM_d(U) \propto U^{-\alpha}dU$ 
where $M_d$ is the dust mass and $U$ is the radiation field intensity and AGN fraction.
The AGN fraction of this component is set to zero since we included another AGN model.
X-CIGALE adopts an energy balance principle for the emission from stars, i.e., the energy emitted
by dust in the IR corresponds to the energy absorbed by dust in the UV to optical.

We adopted models of rest-frame UV to IR emission from an AGN according to \citet{2006MNRAS.366..767F}. 
Here we fixed the parameters 
at default values of X-CIGALE except for position angle ($\psi$) and AGN fraction. 
The parameters fixed here can affect the mid-IR (MIR) SED shape
but this is out of the scope of this study since the MIR flux of our target is not well constrained. 
The $X$-ray power law is parameterized by a photon index $\Gamma$. 
The \citet{2006MNRAS.366..767F} and $X$-ray models are constrained not to 
have an optical to $X$-ray luminosity ratio significantly different 
from the empirical values \citep{2007ApJ...665.1004J}. 
The radio synchrotron emission in X-CIGALE is parameterized with
a power-law spectral slope $\alpha$ where radio flux $F_\nu \propto \nu^{-\alpha}$,
with radio-IR correlation $q_{\rm IR}$ \citep{1985ApJ...298L...7H}. 

We did not include the nebular emission in X-CIGALE 
that is useful for modeling young star-forming galaxies
but not important for a massive quiescent galaxy like our target.
Our target has strong nebular emission lines from an AGN, which is not included in the AGN model templates,
and their contributions to the SED were subtracted in advance.
Low-mass $X$-ray binaries (LMXB), high mass $X$-ray binaries (HMXB), 
and hot gas from a host galaxy were also considered for the $X$-ray model, 
but according to the recipe adopted in X-CIGALE, their contribution to $X$-ray luminosities 
are expected to be two or more orders of magnitude lower than that of the AGN $X$-ray luminosity of our target.
As shown in Figure \ref{fig:spectra}, since it is hard to determine the systemic redshift of the host galaxy, 
hereafter we adopt $z=3.085$ to calculate the physical properties.
We also ran X-CIGALE with a fixed position angle $\psi=0$ deg,
where the contribution of the AGN is negligible at rest-frame UV to optical wavelength. 

\subsection{Emission lines fitting}\label{subsec:spectralfit}

Figure \ref{fig:spectra} shows the [O {\footnotesize II}], H$\beta$,
and [O {\footnotesize III}] spectra of J221737.29+001823.4. 
All the [O {\footnotesize III}] line profiles have significantly redshifted tails. 
H$\beta$ shows no significant emission from a broad line region (FWHM $>2000$ km s$^{-1}$).
We measured the fluxes and line profiles of the emission lines by fitting them with combinations of Gaussian profiles.
The composites of two or three Gaussian profiles are often used to fit the [O {\footnotesize III}] line profiles of AGNs 
(e.g., \citealt{2014MNRAS.441.3306H, 2014MNRAS.442..784Z}). 
\citet{2014MNRAS.442..784Z} found that three-components Gaussian models are better than two or one-component Gaussian models 
for 400 out of 568 obscured AGNs found in SDSS. 
According to \citet{2011MNRAS.418.2032V}, such a three-component Gaussian model
consists of two narrow components 
and one very broad (FWHM $\gtrsim 1000$ km s$^{-1}$) component.
Such a broad component can appear as a distinct tail in a spectrum
but it was not identified well in our target maybe due to the low sensitivity. 
As we describe in later, the line profiles of our target are approximated well 
by combinations of the two narrow Gaussian profiles without another very broad component. 
Thus to compare the line properties properly with literatures, 
we fitted each [O {\footnotesize III}] by with a combination of two Gaussian components. 

The line widths, central wavelengths, and intensity of the two Gaussians are free parameters. 
[O {\footnotesize III}]$\lambda 4959$ was fitted simultaneously with [O {\footnotesize III}]$\lambda 5007$, 
adopting the same model profile and a fixed line ratio [O {\footnotesize III}]$\lambda 4959$/[O {\footnotesize III}]$\lambda 5007$ = 0.33. 
First, we subtracted the residual sky from each spectrum. 
The spectra significantly affected by OH airglow, 
which were automatically defined as regions where $\rm\sigma>1-5\times10^{-17}$ erg cm$^{-2}$ s$^{-1}$~\AA$^{-1}$,
were masked in advance (gray shaded in Figure \ref{fig:spectra}) and not used in the fits.
We then fitted the above models with MCMC using {\sf emcee} \citep{2013PASP..125..306F} in Python.
The 5th, 50th, and 95th percentiles of the samples 
in the marginalized distributions are quoted as means and uncertainties. 
The line center is defined as velocity $v_{50}$ at which 50 percent of the line flux accumulates. 
The line width is defined as $W_{80}$, the velocity width of the line containing 80\% 
of the total flux ($W_{80}=v_{90}-v_{10}$; see details in \citealt{2014MNRAS.441.3306H}). 
We also measured a dimensionless relative asymmetry $=((v_{95}-v_{50})-(v_{50}-v_{05}))/(v_{95}-v_{05})$ \citep{2014MNRAS.442..784Z}.

The fitting procedure for H$\beta$ and [O {\footnotesize II}]
was similar to that of [O {\footnotesize III}], but adopted different models.  
H$\beta$ was fitted with the best-fit model profiles for [O {\footnotesize III}] for each slit
multiplied with a free parameter.
The [O {\footnotesize II}] doublet was fitted using a combination of two Gaussian profiles 
with the same line widths and fixed separation of the central wavelengths 
but with the line ratio as a free parameter. 
Finally, we corrected their instrumental broadening for $W_{\rm 80}$ 
assuming $W_{\rm 80, corr}=[{W_{\rm 80}}^2 - {W_{\rm 80, inst}}^2]^{1/2}$
where the $W_{\rm 80}$ is the original value obtained in the spectral fitting and $W_{\rm 80, inst}$ 
is $W_{80}$ for a Gaussian with the width owed to the instrumental broadening $\approx 60$ km s$^{-1}$.
The corrections are less than 2\% of the original values. 
We evaluated the spatial extent of the  [O {\footnotesize III}] along the slits
by taking the average spatial profiles between 20420 and 20490 \AA.

Figure \ref{fig:spectra} shows the best-fit spectra and 
spatial extent of the [O {\footnotesize III}] emission line at each position. 
The red curves show the best-fit profiles. 
The best-fit model profiles for [O {\footnotesize III}] give reasonable fits for H$\beta$ 
while the observed [O {\footnotesize II}] may trace only a blue part of the [O {\footnotesize III}] 
as the green vertical lines in Figure \ref{fig:spectra} show the $z_{50}$ of [O {\footnotesize II}] corresponding to the [O {\footnotesize III}].

\begin{figure*}
\gridline{\fig{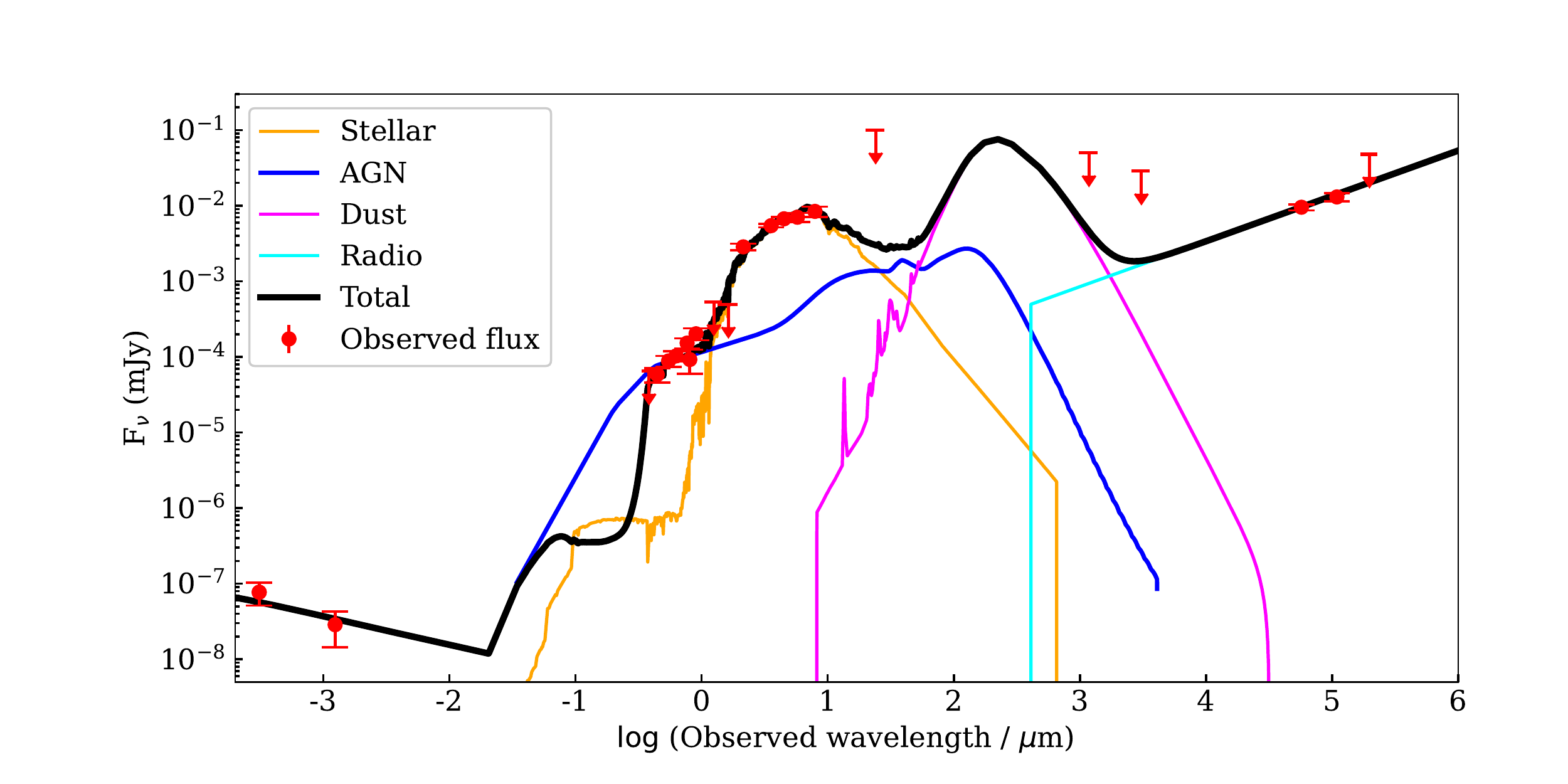}{0.7\textwidth}{}}
\caption{ The best-fit SED of the target obtained with X-CIGALE. 
The red points show the observed fluxes with 1 $\sigma$ errors. 
The allows show the upper limit values. 
The thick black curve shows the best-fit SED that is a composite
of stellar emission (magenta), continuum emission from AGN (blue), dust emission (magenta), and radio synchrotron emission (cyan).
The individual model components are not corrected for absorption by the intergalactic medium.
\label{fig:bestfit}}
\end{figure*}

\begin{deluxetable*}{llcc}
\tablenum{3}
\tablecaption{SED parameters \label{tab:sed}}
\tablewidth{0pt}
\tablehead{
\colhead{Model} & \colhead{parameter}& \colhead{value (free $\psi$)$^a$} & \colhead{value ($\psi = 0$)$^b$}}
\decimalcolnumbers
\startdata
 &  Reduced $\chi^2$  &    0.91 &    1.60 \\
Stellar    & age$_{\rm main}$ / Gyr & $1.68\pm0.29$& $1.40\pm0.04$ \\
               & $\tau_{\rm main}$ / Myr & $45\pm54$ & $200\pm0$ \\
               & $E(B-V)$ / mag & $0.12\pm0.04$ & $0.09\pm0.02$ \\
               & age$_{\star}$$^c$ / Gyr & $1.59\pm0.29$ & $1.00\pm0.36$ \\
               & $M_{\star}$ /  $10^{10}$ M$_{\odot}$ & $9.91\pm1.67$ & $6.22\pm0.33$ \\
               & $SFR$ / M$_{\odot}$ yr$^{-1}$ & $0.05\pm0.02$ & $3.41\pm0.84$\\
Dust       & $\alpha$  & $0.71\pm0.70$ & $0.14\pm0.11$  \\
Fritz06   & $\psi$ / deg  & $51\pm3$ & 0  \\
                & $f_{\rm AGN}$ & $0.16\pm0.16$ & $0.14\pm0.14$ \\
$X$-ray  & $\Gamma$ & $0.67\pm0.20$ & $0.65\pm0.21$  \\
               & $\log (L_{\rm 2-10 keV}$ / erg s$^{-1}$) & $43.2\pm0.1$ & $43.2\pm0.1$ \\
Radio     & $q_{\rm IR}$ &  $0.61\pm0.28$ &  $0.61\pm0.27$  \\
               & $\alpha$  & $0.61\pm0.18$ & $0.73\pm0.22$ \\
               & $\log (L_{\rm 1.4 GHz}$ / erg s$^{-1}$ Hz$^{-1})$ & $31.05\pm0.06$ & $31.08\pm0.09$  \\
\enddata
\tablecomments{$^{(a)}$The values and errors calculated with X-CIGALE. \\
$^{(b)}$Calculated with a fixed position angle $\psi$ = 0 deg.\\
$^{(c)}$The stellar mass weighted age. \\}
\end{deluxetable*}

\begin{deluxetable*}{ccccccc}
\tablenum{4}
\tablecaption{Detected emisson lines\label{tab:lines}}
\tablewidth{0pt}
\tablehead{
\colhead{Slit} & \colhead{line} & \colhead{$z_{50}$} & \colhead{$W_{80, corr}$} & Rel. asym.  & \colhead{Flux}  &  [O{\footnotesize II}]$\lambda3727/ \lambda3729$ \\
\colhead{} & \colhead{} & \nocolhead{} & \colhead{(km s$^{-1}$)} & &\colhead{($\rm 10^{-17}~erg~cm^{-2}~s^{-1}$)} & \colhead{} 
}
\decimalcolnumbers
\startdata
A & [O{\footnotesize III}]$\lambda5007$ & $3.0850_{-0.0002}^{+0.0002}$ & $1174^{+133}_{-50}$ & $0.14^{+0.03}_{-0.03}$ & $21.0_{-0.8}^{+0.8}$ & ... \\
   & H$\beta$                                                  &                                                        &                                       &                                          & $2.2_{-0.2}^{+0.2}$ &  ... \\
   & [O{\footnotesize II}]$\lambda\lambda3727,3729$ & $3.0821_{-0.0010}^{+0.0012}$ & $354_{-128}^{+128}$ & & $1.4_{-0.3}^{+0.3}$ &  $1.0_{-0.3}^{+1.8}$ \\
B & [O{\footnotesize III}]$\lambda5007$ & $3.0851_{-0.0002}^{+0.0002}$ & $1148_{-47}^{+140}$ & $0.15^{+0.04}_{-0.03}$ & $22.7_{-0.9}^{+0.9}$  & ...  \\
   & H$\beta$                                                  &                                                        &                                       &                                          & $2.0_{-0.2}^{+0.2}$ & ... \\
C & [O{\footnotesize III}]$\lambda5007$ & $3.0857_{-0.0005}^{+0.0006}$ & $903_{-91}^{+254}$ & $0.07^{+0.11}_{-0.07}$ & $19.9_{-2.1}^{+2.6}$  & ... \\
   & H$\beta$                                                  &                                                        &                                       &                                          & $2.0_{-0.5}^{+0.5}$ & ...  \\
\enddata
\end{deluxetable*}

\begin{deluxetable}{ccccc}
\tablenum{5}
\tablecaption{The [O {\footnotesize III}] line profile \label{tab:lineprofiles}}
\tablewidth{0pt}
\tablehead{
\colhead{Slit} & z& \colhead{${V_{\rm offset}}^{a}$} & \colhead{FWHM$_{corr}$} & \colhead{fraction$^{b}$}  \\
\colhead{} & \colhead{} & \colhead{(km s$^{-1}$)} & \colhead{(km s$^{-1}$)} & \colhead{} 
}
\decimalcolnumbers
\startdata
A & $3.0832_{-0.0001}^{+0.0002}$ & ...                              &  $242^{+24}_{-17}$ & $0.68_{-0.06}^{+0.04}$ \\
    & $3.0905_{-0.0036}^{+0.0005}$ & $533_{-265}^{+36}$ &  $253^{+83}_{-24}$ & $0.32_{-0.04}^{+0.06}$ \\
B & $3.0832_{-0.0001}^{+0.0002}$ & ...                             &  $229_{-17}^{+25}$ & $0.67_{-0.07}^{+0.04}$ \\
    & $3.0905_{-0.0027}^{+0.0005}$ & $535_{-245}^{+32}$ &  $240_{-21}^{+79}$ & $0.33_{-0.04}^{+0.07}$ \\
C & $3.0848_{-0.0005}^{+0.0005}$ & ...                             &  $147_{-54}^{+69}$ & $0.81_{-0.22}^{+0.08}$ \\
    & $3.0914_{-0.0037}^{+0.0009}$ & $484_{-247}^{+73}$ &  $\lesssim180$ & $0.19_{-0.08}^{+0.21}$ \\
\enddata
\tablecomments{$^{(a)}$ The velocity offset of this component from the brighter component.\\
$^{(b)}$ The fraction of this component in the total flux.\\
 }
\end{deluxetable}

\section{Result} \label{sec:result}

\subsection{SED of the host galaxy} \label{subsec:sedhost}

Figure \ref{fig:bestfit} and Table \ref{tab:sed} show the best-fit SED and parameters of J221737.29+001823.4, respectively. 
The AGN dominates the rest-frame UV emission.
Similar to our previous SED modeling without an AGN 
component (\hyperlink{2015ApJ...799...38K}{K15}), 
the best-fit SED is that of a massive quiescent galaxy 
but due to the contribution of the AGN to the rest-frame UV, 
old stars are more dominant where the best-fit age and stellar mass become larger. 
The reduced $\chi^2$ value for the $\psi = 0$ model is larger than that for the $\psi$ free model 
but still, the $\psi=0$ model can fit the observed fluxes well. 
Therefore, we cannot reject the possibility that J221737.29+001823.4 has a smaller $\psi$ in fact
and the AGN contribution on the SED at rest-frame UV is overestimated. 

The best-fit SED from X-CIGALE at IR is lower than the upper limit of the observed fluxes.
Adopting the average 1.2 mm flux to IR luminosity $L_{IR}$ relation for the SED library in \citet{2017ApJ...840...78D}, 
the conservative upper limit of $L_{IR}$ is $\sim0.9-2.0\times10^{11}~L_{\odot}$ taking the 95\% confidence interval.
It corresponds to SFR $\rm<9-21~M_{\odot}$ yr$^{-1}$ using the $L_{\rm IR}$ to SFR 
conversion in \citet{2012ARA&A..50..531K}. 
Hereafter we use $L_{IR}=2.0\times10^{11}~L_{\odot}$ 
and SFR $=21~M_{\odot}$ yr$^{-1}$ to estimate a conservative upper limit. 

\subsection{AGN luminosity} \label{subsec:linestrength}

Table \ref{tab:lines} lists the strength, line center and line width at each slit position. 
Using the largest flux value at slit-B, J221737.29+001823.4 has a [O {\footnotesize III}] 
luminosity of $\log (L_{\rm [O III]}\rm~/~erg~s^{-1}) =43.28\pm0.01$. 
The observed $X$-ray luminosity is $\log (L_{\rm2-10 keV}\rm/erg~s^{-1}) = 43.2\pm0.1$.
Thus $L_{\rm [O III]}$ is more than ten times larger than that expected from the empirical 
$X$-ray to  [O {\footnotesize III}]  luminosity relation for type-1 AGNs in the local Universe
(e.g., \citealt{2015ApJ...815....1U}). 
Furthermore, our target has a flat $X$-ray SED ($\Gamma = 0.67\pm0.20$) that suggests significant absorption on $X$-ray
where $\Gamma\sim1.8$ in typical (e.g., \citealt{2016ApJ...831..145Y,2017ApJS..232....8L}). 
We then estimated the intrinsic $X$-ray luminosity based on the [O {\footnotesize III}] luminosity.  
Adopting the $X$-ray to [O {\footnotesize III}] luminosity relation 
for Seyfert-1 AGNs in the local Universe \citep{2015ApJ...815....1U}, 
J221737.29+001823.4 has an intrinsic $X$-ray luminosity of $\log (L_{\rm2-10 keV, intrinsic}\rm/erg~s^{-1}) \approx 44.4$. 
Applying the $X$-ray to bolometric correction as in \citet{2004MNRAS.351..169M}, 
the source has a bolometric luminosity of $\log (L_{\rm bol}\rm/erg~s^{-1}) \approx 46.0$, 
which would make it a moderately luminous QSO. 
Assuming a scaling relation between the black hole mass
and the host galaxy stellar mass of  $M_{\rm BH} \approx 0.002~M_{\star}$,
following \citet{2012ApJ...746...90A},
J221737.29+001823.4 may have $M_{\rm BH} \approx 2\times10^8~M_{\odot}$
and Eddington luminosity $L_{\rm Edd} = 1.3\times10^{38}~M_{\rm BH}/M_{\odot} \approx 2.6\times10^{46}$ erg s$^{-1}$.
Thus it may have a relatively large Eddington ratio of $L_{\rm bol}/L_{\rm Edd} \sim0.4$.

\subsection{Radio emission} \label{subsec:radio} 

J221737.29+001823.4 has a radio spectral index $\alpha=0.61\pm0.18$ where $F_\nu \propto \nu^{-\alpha}$,
which is a typical value of radio synchrotron emission (e.g., \citealt{2017A&A...602A...1S}).
The luminosity at 1.4 GHz is given as $L_{\rm 1.4 GHz} = F_{\rm1.4 GHz} 4\pi {D_L}^2 (1+z)^{(\alpha -1)} $ 
where $D_L$ is a luminosity distance, and $F_{\rm1.4 GHz}$ is the flux at 1.4 GHz, which is $0.021\pm0.003$ mJy from X-CIGALE. 
Then we find a radio luminosity of $\log (L_{\rm 1.4 GHz}$ $\rm/erg~s^{-1}~Hz^{-1}) = 31.05\pm0.06$. 
It would be classified as a radio-quiet AGN adopting the criteria suggested by \citet{1999AJ....118.1169X}. 
The logarithmic ratio of the infrared and radio fluxes $q_{\rm IR}$ of J221737.29+001823.4 is $0.61\pm0.28$ for the best-fit model of X-CIGALE 
where $q_{\rm IR}$ is defined as $\log_{10} (F_{\rm FIR}/(3.75\times10^{12} Hz)/F_{\rm 1.4 GHz})$ \citep{1985ApJ...298L...7H}. 
Using a conservative upper limit on $L_{\rm IR}$, $q_{\rm IR}$ is $\lesssim1.4$.
It is lower than that of pure star-forming galaxies with 2.4 in \citet{2010A&A...518L..31I} 
and comparable to radio-loud AGNs at low redshift (e.g., \citealt{2019ApJS..243...15T}).
At high redshift, QSOs selectable by wide-field surveys like SDSS have $q_{\rm IR}$ similar to star-forming galaxies \citep{2008MNRAS.386..953I,2009ApJ...706..482M}
while radio sources with low $q_{\rm IR}$ like local radio-loud AGNs are found by deep surveys like VLA COSMOS 3 GHz survey \citep{2017A&A...602A...4D}.
Thus J221737.29+001823.4 is one of the most local radio-loud AGN-like objects selectable by the radio surveys to date.

\begin{figure}
\gridline{\fig{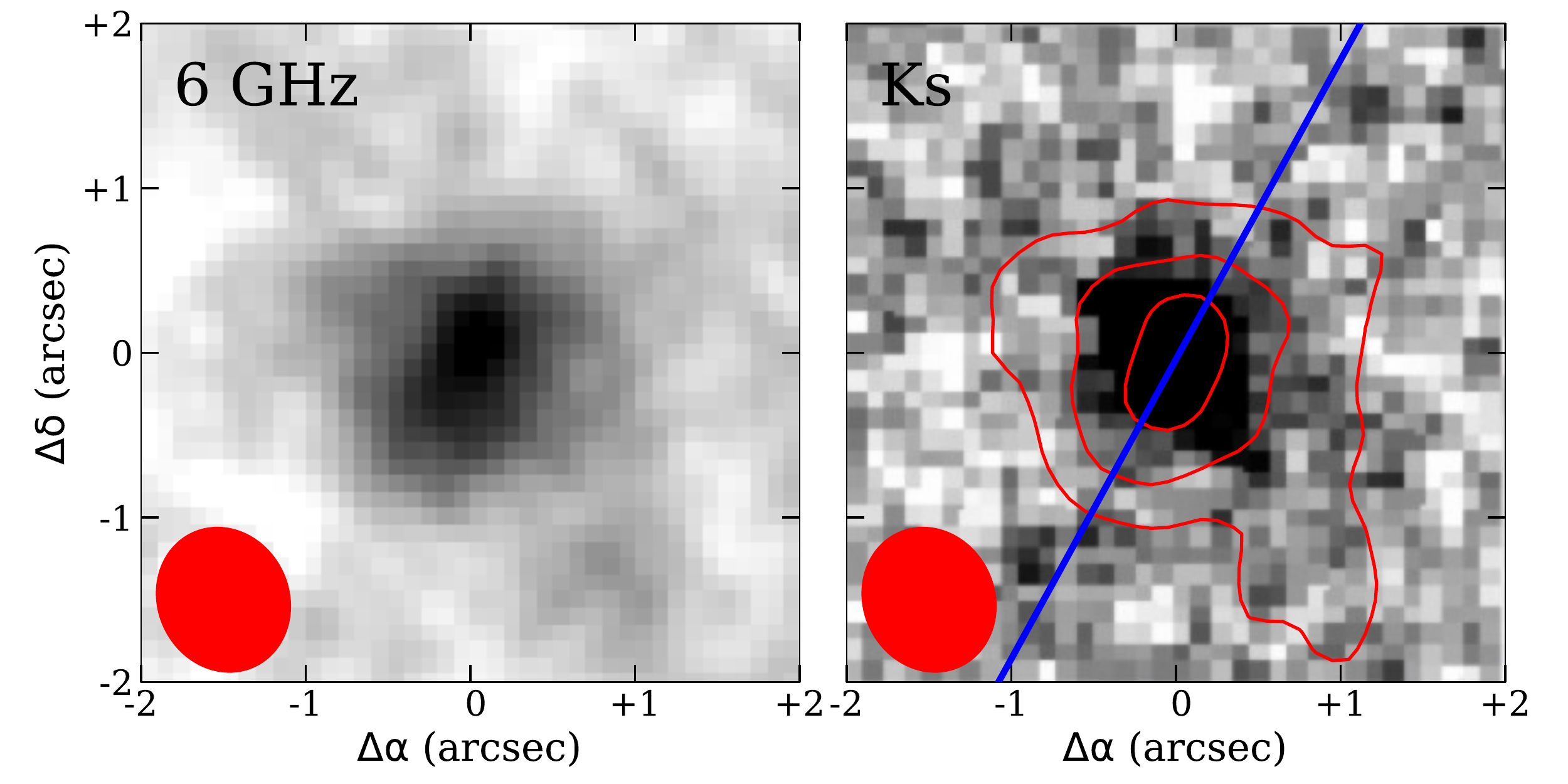}{0.47\textwidth}{}}
\caption{ Left: The 6 GHz image of the target. The image size is 4 arcsec by side. 
The red filled ellipse shows the beam size at 6 GHz. 
Right: The 3, 6, and 9 $\sigma$ contours (red) of the 6 GHz are overlaid on the $K_s$-band image of the target. 
The blue line shows the position angle measured at 6 GHz. 
\label{fig:radioimage}}
\end{figure}

Figure \ref{fig:radioimage} shows the 6 GHz image
and the flux contours of the radio emission overlayed on the $K_s$-band image.
J221737.29+001823.4 is well resolved at 6 GHz 
where the image component size after deconvolution of the beam
are $(0.99\pm0.05) \times (0.52\pm0.05)$ arcsec and PA $=151.2\pm4.6$ deg.
There is also a diffuse component with an orientation different from that of the bright central source. 
The source lies at the low power end of the distribution for compact steep spectrum radio galaxies (e.g., \citealt{1994ApJS...91..491G, 1998PASP..110..493O})
that are believed to be young systems evolving into more extended radio galaxies 
or the systems prevented the spatial growths by the dense interstellar medium (ISM). 
The radio position angle is different from the slit directions for NIR spectroscopy
and then it is not clear whether the [O {\footnotesize III}] traces the radio morphology. 
But the similar spatial extents of [O {\footnotesize III}] and radio emission can indicate 
their physical connection though both are limited by the observational depths.

\subsection{Outflow} \label{subsubsec:outflow}
 
Although the small notches of the spectra left from the model spectra indicate further more complex kinematics, 
each spectrum is well approximated by a combination of two Gaussian profiles.
The obtained spectra have large relative asymmetries $0.07-0.15$ and line widths of $\approx 1000$ km s$^{-1}$.
The spatial extent of [O {\footnotesize III}] ($\approx 15$ kpc in Figure \ref{fig:spectra}) of J221737.29 +001823.4 
is much larger than the typical size of a massive quiescent galaxy at high redshift
(a few physical kpc; e.g., \citealt{2015ApJS..219...15S}) 
though the size of the host galaxy cannot be evaluated robustly since 
the $K_s$-band image can be significantly contaminated by [O {\footnotesize III}] emission.
These line profiles observed at each slit
indicate a significant outflow of ionized gas spread over the host galaxy.
Such galactic wide-scale outflows of ionized gas have been observed among AGNs
in the local Universe and at high redshift
(e.g., \citealt{2014MNRAS.441.3306H,2014ApJ...796....7G,2016MNRAS.456.1195H,2018ApJ...866..119L,2020A&A...642A.147K}).

Table 4 summarizes the ingredient of each spectrum. 
At all slit locations, a spectrum consists of two narrow components.
The redshift of the bluer components at slit-A and B match with that of [O {\footnotesize II}] within errors 
while the redshifts of both the two components at slit-C are redder than [O {\footnotesize II}].
Therefore, J221737.29 +001823.4 consists of one component also detected with [O {\footnotesize II}] at $z\approx3.0832$ (i), 
and at least two redshifted components 
( ii. bluer component at slit-C, and iii. redder components at slit-A, B and C, at the resolution of this study).
Comparing with local AGNs with [O {\footnotesize III}] line width as wide as J221737.29 +001823.4 
in literature \citep{2011MNRAS.418.2032V,2012ApJ...745...67F,2014MNRAS.442..784Z}, 
J221737.29 +001823.4 has components with narrower line width 
and a larger velocity offset between the first and second brightest components. 
Then the wide line widths of the whole line profiles of J221737.29 +001823.4 are due to the large velocity offsets of the two components 
while those are due to broad components for most of the type-2 AGNs.
The broader components found in such type-2 AGNs are generally blueshifted from the narrower components 
and regarded as highly disturbed outflowing gas. 
In the case of J221737.29 +001823.4, the component-(iii) seen at all slit positions likely comes from the gas near the central SMBH. 
Since the emission lines are spatially extended at all slit directions, 
component-(i) and (iii) which dominate the emission-line fluxes at slit-A and B, and slit-C, respectively, 
likely come from the outflowing extended emission-line region. 
The narrower line width indicates that the outflowing gas is not highly perturbed like typical type-2 AGNs with wide line width. 
Maybe the outflowing gas of this system is not disturbed because of young age, lack of intergalactic gas to interact, or the wing of the gas is highly attenuated by dust.

Here we estimate the mass outflow rate of J221737.29 +001823.4 following \citet{2017A&A...601A.143F}, 
using the [O {\footnotesize III}] $W_{80,corr}$ and the spatial extent and H$\beta$ luminosity for slit-A
which are similar to those for slit-B and slit-C.
Based on the H$\beta$ luminosity, the mass of the outflowing ionized gas is estimated as
$$ M_{\rm H\beta} = 7.8 \times10^8 C \left(\frac{L_{H\beta}}{10^{44}}\right)\left(\frac{\langle n_e\rangle}{10^{3}}\right)^{-1},$$
following \citet{2006agna.book.....O} and \citet{2015A&A...580A.102C}, 
where $C=\langle n_e\rangle^2/\langle n_e^2\rangle$ (set to unity
which is a conservative lower limit) and a gas temperature $T=10^4$ K is assumed. 
Adopting $n_e=200$ cm$^{-3}$ following \citet{2017A&A...601A.143F}, 
we find $M_{\rm H\beta} = 7.2^{+0.7}_{-0.7}\times10^7$ M$_{\odot}$.

\begin{figure*}
\gridline{\fig{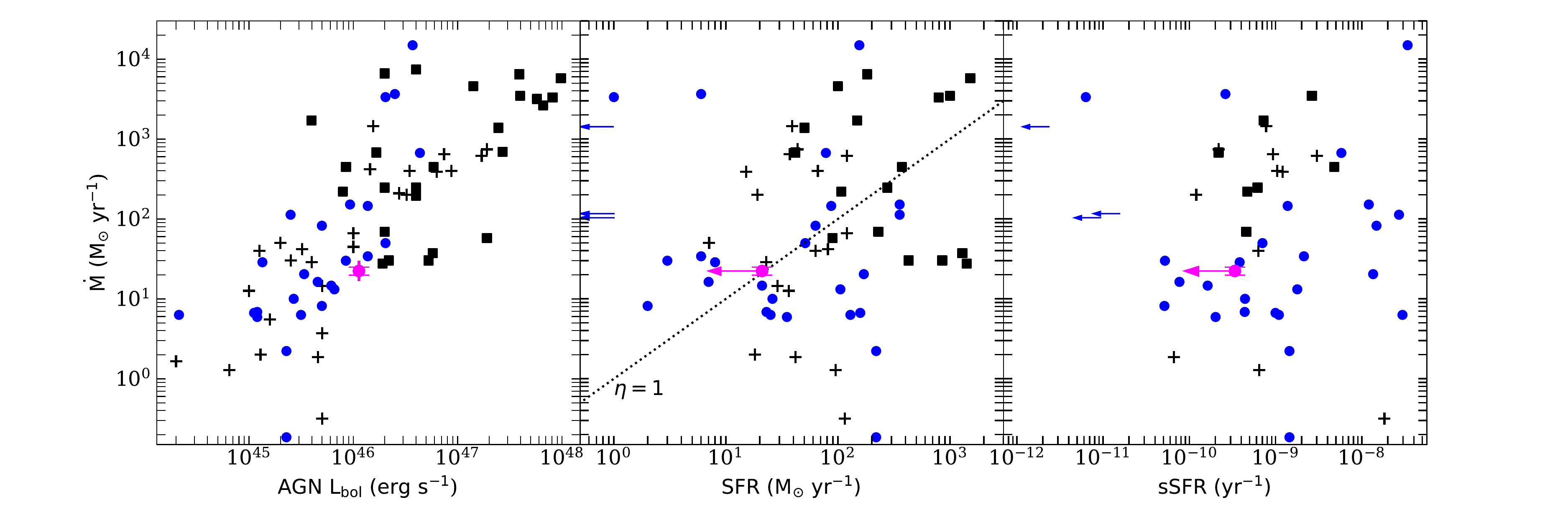}{1.0\textwidth}{}}
\caption{Mass outflow rate vs. AGN bolometic luminosity $L_{\rm bol}$ (left), and SFR (center) and sSFR (right) of the host galaxy.
The filled magenta circle shows J221737.29+001823.4. 
The black crosses and filled squares show the QSOs at low and high redshift from \citet{2017A&A...601A.143F}. 
The blue-filled circles show AGNs in \citet{2019ApJ...886...11L}. 
We corrects the mass outflow rate in \citet{2019ApJ...886...11L} adopting the equation in \citet{2017A&A...601A.143F}. 
The dashed line in the central panel shows $\eta=1$. 
\label{fig:lbol_mdot}}
\end{figure*}

The mass outflow rate of ionized gas is calculated as, 
$$ \dot{M} = 3 \times v_{max} \times M_{\rm H\beta}/R,$$
where $v_{max}$ is the wind maximum velocity, and $R$ is the radius at which the mass outflow rate is computed. 
$v_{max}$ is defined as $W_{80,corr}/1.3$ 
and $R$ is defined as the maximum radius at which the high velocity gas is detected. 
Here we use half of the extent of [O {\footnotesize III}] for each slit, $\approx7.1$ kpc, 
corrected of the spatial resolution with FWHM PSF of $\approx 4.5$ kpc on average.  
We find a mass outflow rate $\rm 22\pm3~M_{\odot}$ yr$^{-1}$. 
Its mass loading factor $\eta = \dot{M}$/SFR is higher than unity, or the $\dot{M}$ exceeds the SFR. 
Note that our estimate does not include outflowing gas at a neutral or molecular state. 
Including them, the total mass outflow rate and thus $\eta$ could be a few to ten times larger than that of the ionized gas 
(e.g., \citealt{2013ApJ...768...75R,2015A&A...580A.102C,2017A&A...601A.143F,2019MNRAS.483.4586F}). 

We compared the mass outflow rate, AGN bolometric luminosity, SFR, and specific SFR (sSFR) of J221737.29+001823.4 with 
results in \citet{2017A&A...601A.143F} 
who summarized outflow properties of QSOs at low and high redshift in the literature,
and \citet{2019ApJ...886...11L} 
who investigated outflow properties for $X$-ray AGNs at $1.4<z<3.8$
in Figure  \ref{fig:lbol_mdot}.
We used the upper limit on SFR from the 1.2 mm flux ($\rm <21~M_{\odot}$ yr$^{-1}$) 
and the stellar mass estimated assuming $\psi=0$ to obtain sSFR as a conservative limit. 
J221737.29+001823.4 is not a special object; 
its ionized gas mass outflow rate is similar 
to those of AGNs with similar bolometric luminosities. 
As previous studies have shown, the SFRs of AGNs with strong outflows 
range widely from quiescent (sSFR $<10^{-10}$ yr$^{-1}$) to starbursting. 

\begin{figure*}
\gridline{\fig{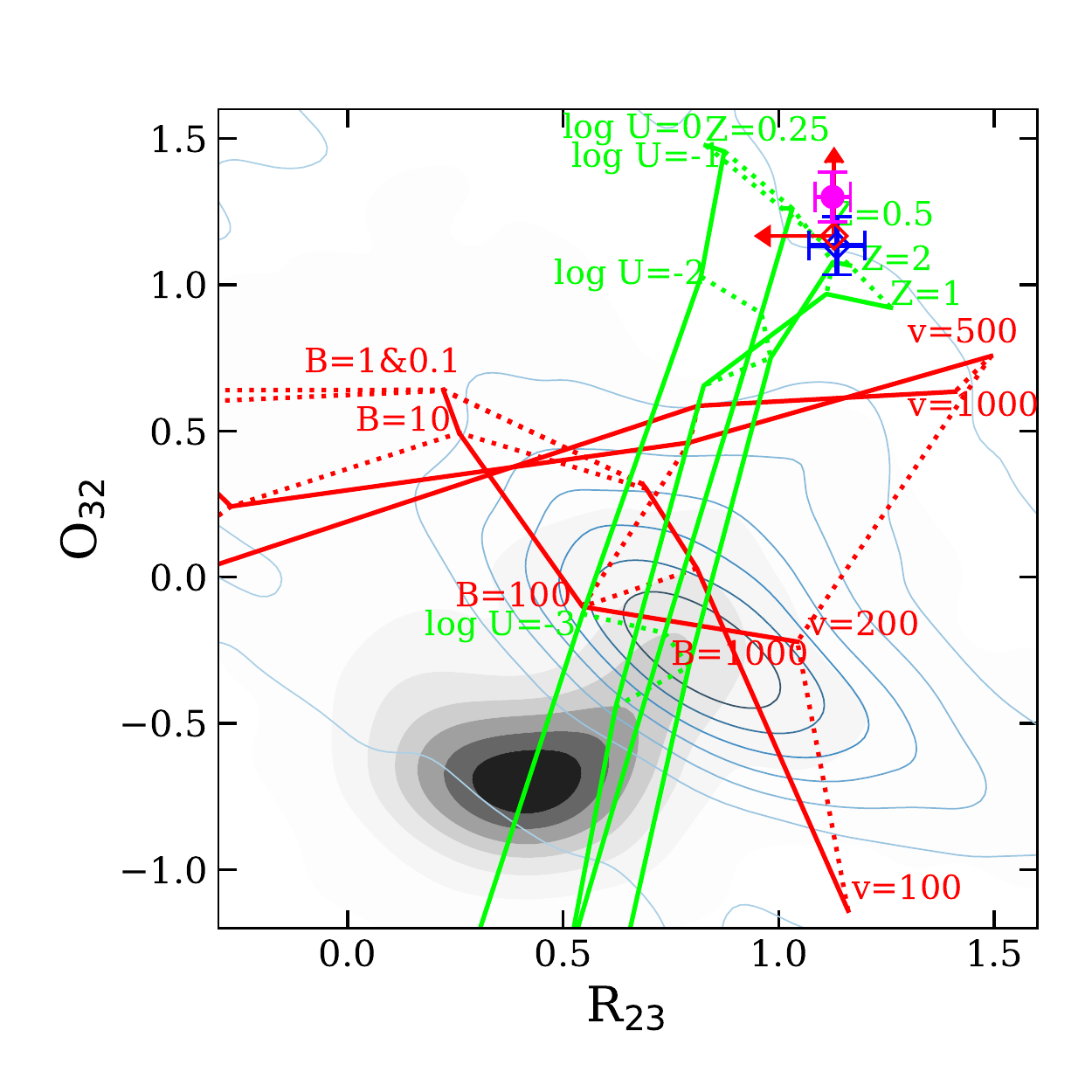}{0.45\textwidth}{}\fig{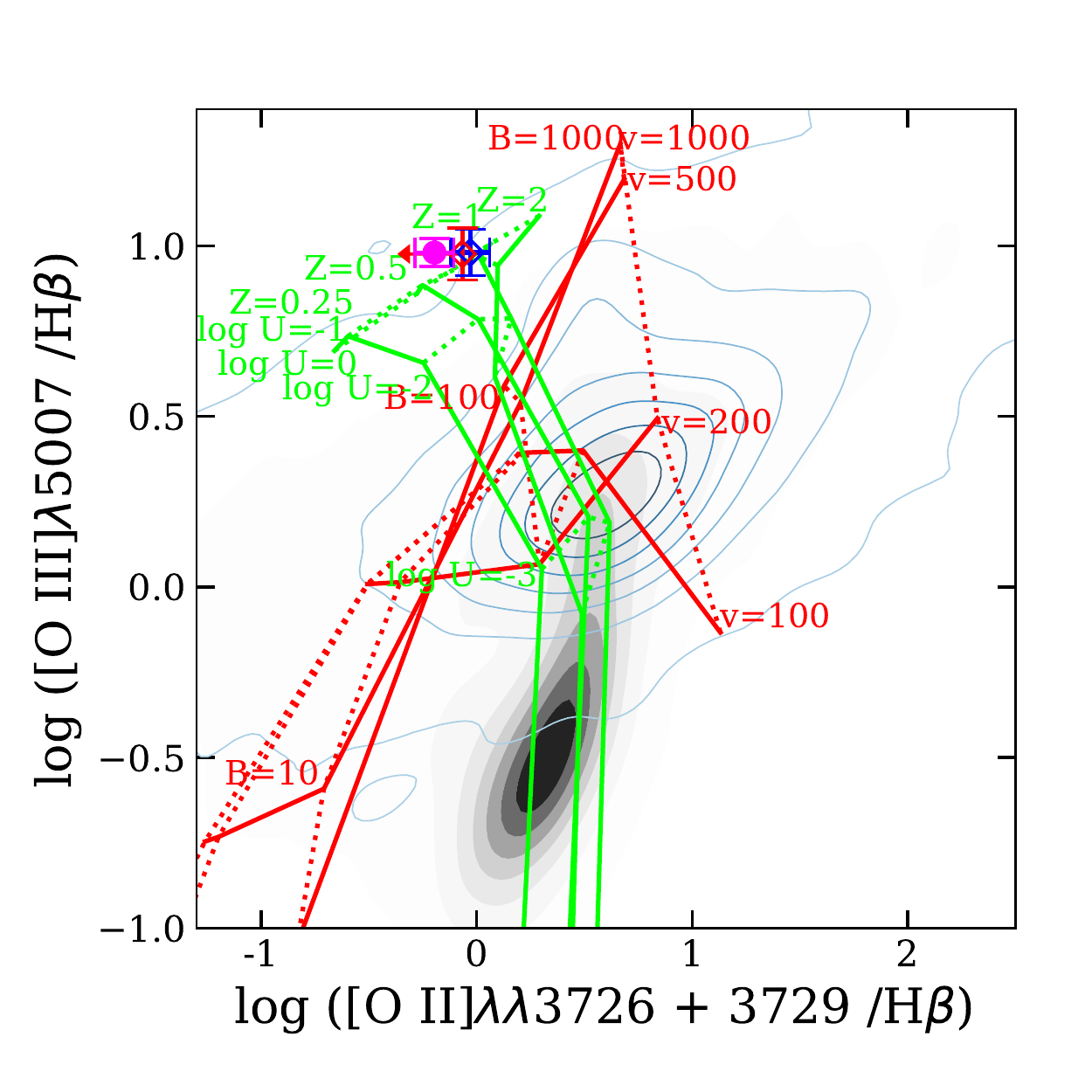}{0.45\textwidth}{}}
\caption{ {\it Left:}  O32 vs. R23 indices. The filled magenta circle with error bars shows the target. 
The filled black contour shows the distribution of star-forming galaxies from SDSS (based on MPA-JHU catalog; \citealt{2004MNRAS.351.1151B,2004MNRAS.353..713K,2004ApJ...613..898T}). 
The blue contour shows the distribution of AGNs selected from SDSS.
Both the star-forming galaxies and AGNs are selected based on BPT diagram.  
The green tracks show the photoionization model from \citet{2004ApJS..153....9G}. 
The models with $\alpha=-2$, $\log U$ = -4, -3, -2, -1, 0, and $Z$ = 0.25, 0.5, 1, \& 2 $Z_{\odot}$ are shown. 
The red tracks show the shock ionization model from \citet{2008ApJS..178...20A}.
The models with $n=1000$ cm$^{-3}$, $Z=Z_{\odot}$, $v$ = 100, 250, 500, \& 1000 km s$^{-1}$, 
and $B$ = 0.1, 10, 100, \& 1000 $\mu$G are shown. 
{\it Right:}  Similar to the left panel but  for [O {\footnotesize III}]$\lambda$5007/H$\beta$ 
vs. [O {\footnotesize III}]$\lambda\lambda$3726+3729/H$\beta$ line diagnostics.
\label{fig:linediagnostics}}
\end{figure*}

\subsection{Emission line diagnostics} \label{subsec:linediagnostics}

We show the emission line diagnostics of J221737.29 +001823.4 measured at slit-A in Figure \ref{fig:linediagnostics}. 
The left panel shows O32 vs. R23, where O32 and R23 are defined as
$\log $ ([O {\footnotesize III}]/[O {\footnotesize II}]) and  
$\log $ (([O {\footnotesize III}]+[O {\footnotesize II}])/H$\beta$), respectively.
The right panel shows [O {\footnotesize III}]$\lambda$5007/H$\beta$ 
vs. [O {\footnotesize II}]$\lambda\lambda$3726,3729/H$\beta$ line diagnostics. 
The magenta filled circle shows the line ratio of J221737.29 +001823.4.
The blue and red diamonds show the line ratios of the bluer and redder components.
The black filled contour and blue contour show star-forming galaxies and AGNs, respectively, 
selected from Sloan digital sky survey (SDSS) 
using spectroscopic data products from the Max Planck Institute for Astrophysics 
and Johns Hopkins University DR8 catalog (MPA-JHU, \citealt{2004MNRAS.351.1151B,2004MNRAS.353..713K,2004ApJ...613..898T}). 
The star-forming galaxies and AGNs are classified using BPT diagrams \citep{1981PASP...93....5B} following \citet{2006MNRAS.372..961K}.  
We applied extinction correction using the H$\alpha$/H$\beta$ line ratio adopting the \citet{1989ApJ...345..245C} extinction law for SDSS 
while no extinction correction was applied for our target. 

J221737.29+001823.4 has large O32 and [O {\footnotesize III}]$\lambda$5007/ H$\beta$ indices 
and is at the edge of the distribution of local AGNs. 
Such line ratios are seen among young star-forming galaxies with high ionization parameters 
at $z=2\sim3$ \citep{2018ApJ...868..117S} 
while AGN is plausibly the dominant ionizing source for our target
given the low SFR and the evolved stellar population.
\citet{2018ApJ...866..119L} finds similarly high [O {\footnotesize III}]$\lambda$5007/H$\beta$
for the optically faint ($R=20-25$) AGNs at $z\sim2-3$
and they are certainly classified as AGNs using BPT diagram.

We compare the line ratios with models of photoionization by AGNs \citep{2004ApJS..153....9G} 
and shock excitation by outflows \citep{2008ApJS..178...20A}.
\citet{2004ApJS..153....9G} provides line intensities predicted for dusty 
and radiation pressure dominated photoionzation models using MAPPING III code \citep{1993ApJS...88..253S}. 
It describes the intensities of the emission lines from the narrow line region (NLR) of AGNs
with power law ionizing SED with hydrogen density $n_{\rm H} = 1000$ cm$^{-3}$, 
$\alpha=-1.2$ to $-2.0$, ionization parameter $\log U=-4$ to $0$, and metallicity $Z=0.25$ to $2~Z_{\odot}$. 
Here we show the dusty model in \citet{2004ApJS..153....9G} 
with $\alpha=-2$, $\log U=-4$ to $0$, and $Z=0.25$ to $2~Z_{\odot}$.
\citet{2008ApJS..178...20A} also uses MAPPING III but provides line intensities predicted for shock excitation. 
They provide models with preshock density $n=0.01$ to 1000 cm$^{-3}$, 
velocity $v<1000$ km s$^{-1}$, magnetic field $B=10^{-10}$ to $10^{-4}$ G, 
and magnetic parameter $B/n^{1/2}=10^{-4}$ to 100 $\mu$G cm$^{3/2}$. 
We show the models with $n=1000$ cm$^{-3}$, $v=100$ to $1000$ km s$^{-1}$, and $B=0.1$ to $1000~\mu$G.
We choose $\alpha=-2$ and $n=1000$ cm$^{-3}$ models 
since only these models have line ratios close to our target. 

The line ratios of J221737.29+001823.4 are close to those of the photoionization models with high ionization parameter $\log U>-1$.
It indicates that most of its emission originates  
in the central AGN rather than the shock induced by e.g., radio jets. 
The line ratios of individual velocity components are also consistent with the photoionization models.
Thus the outflowing gas of J221737.29+001823.4 is likely ionized by the central AGN rather than shock excitation.  
It is consistent with the spatially resolved spectroscopic analysis in e.g., \citet{2017ApJ...834...30F,2018ApJ...856...46R} for local AGNs.
We note that there are several known uncertainties in our results.
First, the line ratio based on the slit spectroscopy can be a composite of several independent components. 
For a more detailed discussion, we need a spatially resolved spectroscopy. 
Second, the dust attenuation is not corrected for our target but
is negligible assuming the absorption on emission lines
$E(B-V)_{\rm line} = E(B-V)/0.44 \approx0.2$ measured with X-CIGALE; 
R23 and [O {\footnotesize III}]/H$\beta$ do not change significantly while
O32 can decrease with $\sim0.1$ and log [O {\footnotesize II}]/H$\beta$
can increase with $\sim0.1$ adopting \citet{1989ApJ...345..245C} extinction law. 

We can also measure the electron density ($n_e$) using the [O {\footnotesize II}]$\lambda$3727
to $\lambda$3729 line ratio \citep{2006agna.book.....O}. 
Following \citet{2016ApJ...816...23S}, the line ratio of our target corresponds to 
an electron density $n_e = 10-1500$ cm$^{-3}$, which is within the expected range 
for NLR, $10-10000$ cm$^{-3}$ 
(e.g., \citealt{2006ApJ...650..693N,2013MNRAS.436.2576L}). 

\section{Discussion} \label{sec:discussion} 

We identified a QSO in a quiescent galaxy that plausibly stopped star formation several hundred Myr ago
which is analogous to so-called post-starburst galaxies
selected with Balmer absorption features caused by A-type stars in the local Universe 
though our target is hardly detected of Balmer absorption features with the current facilities. 
Post-starburst galaxies often show significant emission lines those likely originated in AGN rather than star formation
(e.g., \citealt{2006ApJ...648..281Y,2007MNRAS.382.1415S,2010MNRAS.405..933W,2016ApJS..224...38A}).
Similarly, massive quiescent galaxies at high redshift sometimes show weak [O{\footnotesize II}] or [O{\footnotesize III}] 
which are thought to originate in AGNs 
(e.g., \citealt{2010ApJ...716..970L,2018A&A...618A..85S,2020ApJ...905...40S}; \hyperlink{2021ApJ...919....6K}{K21}). 
However, the QSOs with [O {\footnotesize III}] as luminous as our target are rarely found in the local Universe \citep{2019MNRAS.485.2710J}.
The QSOs hosted by quiescent galaxies at high redshift have not been studied in detail previously, 
but they may not be so unusual given the presence of QSOs with low sSFR and strong outflows
(e.g., \citealt{2017A&A...601A.143F,2019ApJ...886...11L}).

In the case of J221737.29+001823.4, photoionization by AGN is likely the dominant excitation mechanism 
but shocks can also be the dominant excitation mechanism of strong emission lines of post-starburst galaxies \citep{2016ApJS..224...38A}. 
The type-2 QSOs with broad and luminous [O {\footnotesize III}] at $z<0.2$ in \citet{2014MNRAS.441.3306H}
support the shock excitation scenario though they share several properties with J221737.29+001823.4;  
they have $\log (L_{\rm [OIII]}$ / erg s$^{-1}) =42-43.2$, [O {\footnotesize III}] line FWHM $=800-1800$ km s$^{-1}$, 
$\log (L_{\rm 1.4 GHz}$ / W erg$^{-1}$ s$^{-1}) = 30.3-31.4$, compact radio sizes (1-25 kpc), 
small $q_{\rm IR}=1-2$, and low SFR of $\lesssim50$ M$_{\odot}$ yr$^{-1}$, and thus are classified as radio-quiet QSOs. 
\citet{2019MNRAS.485.2710J} suggested a scenario in which low-power radio jets
are confined by the ISM for a long time and efficiently affect the ISM over a large volume
and result in strongly ionized gas outflows (e.g., \citealt{2016MNRAS.461..967M}).

How to power a QSO in a quiescent galaxy like J221737.29+001823.4?
Since radio-quiet QSOs are generally star-forming (e.g., \citealt{2012MNRAS.421.1569B, 2019NatAs...3..387P}),
it is not surprising that there is abundant gas to fuel central SMBHs. 
However, J221737.29+001823.4 has likely been quenched for several hundred Myr.
One interesting issue is the origin of the gas supply to power a QSO.
Since J221737.29+001823.4 is at high redshift, cold gas supply can be still abundant
though star formation should be kept quenching according to the observed SED. 

Another possible supply of gas to power a QSO is mass loss from evolved stars following a starburst
\citep{1988ApJ...332..124N,2007ApJ...665.1038C,2009MNRAS.397..135K}. 
In this scenario, the mass loss from asymptotic giant branch 
(AGB) stars can result in a rise of AGN at $100\sim300$ Myr after the starburst
and may explain some observed time-delay of AGNs found in the local Universe 
(e.g., \citealt{2007ApJ...671.1388D, 2009MNRAS.397..135K, 2010MNRAS.405..933W}).
In this scenario, the mass loss from supernova explosions (SNe) supplies gas at early times
during which the mass accretion to central SMBHs is not efficient
because the velocity of the gas ejected by SNe is several thousand km s$^{-1}$. 
At late times, mass loss from AGB stars becomes dominant. 
Since the speed of the gas ejected as planetary nebulae is several ten km s$^{-1}$, 
they can more easily accrete into the central SMBH. 

Since the SFH of massive quiescent galaxies at high redshift is believed to be more bursty than the local star-forming galaxies
(e.g., \citealt{2018A&A...618A..85S,2021ApJ...919....6K,2020ApJ...903...47F, 2020ApJ...889...93V,2020ApJ...905...40S}), 
after the starburst, a large amount of gas should be supplied from evolved stars in a short timescale. 
We calculated the gas production from the mass loss of stars in J221737.29 +001823.4 using {\sf GALAXEV} \citep{2003MNRAS.344.1000B}
assuming a constant starburst for 50 Myr which forms a stellar mass of $5\times{10}^{10}$ M$_{\odot}$  
or half the stellar mass of J221737.29 +001823.4 where the lifetime of SMGs are found $40-200$ Myr 
\citep{2005MNRAS.359.1165G,2006ApJ...640..228T,2014ApJ...782...68T}. 
There is a supply of $\rm\gtrsim 10~M_{\odot} yr^{-1}$ gas 
from mass loss of stars at all times within $\sim1$ Gyr from the beginning of the starburst. 
Using the equation adopted by \citet{2010MNRAS.405..933W}, 
the black hole accretion rate (BHAR) of our target 
is estimated from the [O {\footnotesize III}] luminosity as 
$\rm 4\times10^{-10}\times L_{\rm [OIII]}/L_{\odot} \approx 2~M_{\odot} yr^{-1}$ 
that is reduced if a part of [O {\footnotesize III}] is induced by radio jets. 
It is the 20 \% of the mass loss from stars expected for J221737.29 +001823.4.
We note that \citet{2010ApJ...717..708C} predicted that 2\% of the recycled gas is used to fuel the central SMBH.
From the above, a substantial amount of AGN luminosity can be fueled naturally by mass loss from stars
but we cannot reject the possibility of an external supply of the gas from the circum/intergalactic media to fuel the central SMBH. 

\section{Conclusion} \label{sec:conclusion} 

We have identified a massive quiescent galaxy hosting a QSO with ionized gas outflows
by detecting redshifted [O {\footnotesize III}] $\lambda\lambda$4959,5007 emission lines. 
Large line widths ($W_{80}>1000~\rm km s^{-1}$) measured at the three slit positions
indicate the presence of powerful outflow in multiple directions.
Given the quiescent SED of the host galaxy, 
we conclude that we are witnessing a QSO arising several hundred Myr after the quenching of the starburst. 
According to the emission line diagnostics,
most of the emission line flux is likely results from radiative heating by the QSO rather than radio jets. 

Our results suggest a new aspect of the role of AGNs in galaxy evolution at high redshift.
In one popular galaxy and SMBH co-evolution scenario,
it is believed that a central SMBH is grown with the star formation of a host galaxy, 
QSO-driven outflows remove gas from a host galaxy and quench its star formation, 
after which the accretion rates onto the central SMBHs decline and radio jets regulate the star formation of a host galaxy.  
Indeed there are dusty starburst galaxies and obscured AGNs 
which are believed to be the previous steps of QSOs in this $z=3.09$ protocluster. 
On the other hand, in this case, we find that a QSO appears several hundred Myr after the quenching of star formation.
The presence of powerful outflow suggests that there remains a significant supply of gas to power a QSO after quenching. 
Taken together, it is possible that the progenitors of giant ellipticals may become strong QSOs both before and after quenching. 
We suggest a scenario in which once a QSO and/or stellar feedback initially quenches star formation, 
subsequent accretion onto the SMBH of residual gas can power a QSO effectively, 
and this late-time QSO further completes a quenching.

\acknowledgements

This work has been supported by JSPS KAKENHI Grant Numbers 
20K14530, 21H044902 (MK), 17K14252, 20H01953 (HU), 19H00697, 20H01949 (TN), 17H06130 (YT and KK).
This work has been also supported by a US National Science Foundation (NSF) grant AST-2009278 (CCS).
This work has been also supported by NAOJ ALMA Scientific Research Grant Numbers 2018-09B (YT)
The spectroscopic data were obtained at the W. M. Keck Observatory, 
which is operated as a scientific partnership among the California Institute of Technology, 
the University of California, and the National Aeronautics and Space Administration.
The observations were carried out within the framework of Subaru-Keck/Subaru-Gemini 
time exchange program which is operated by the National Astronomical Observatory of Japan.
The $K_s$-band image was collected with nuMOIRCS at Subaru Telescope, 
which is operated by the National Astronomical Observatory of Japan.
We are honored and grateful for the opportunity of observing the 
Universe from Maunakea, which has the cultural, historical, and natural
significance in Hawaii.
This paper makes use of the following ALMA data: ADS/JAO.ALMA\#2013.1.00162.S, 
ADS/JAO.ALMA\#2016.1.00580.S,
ADS/JAO.ALMA \#2017.1.01332.S. 
ALMA is a partnership of ESO (representing its member states), NSF (USA), and NINS (Japan), together with NRC (Canada), MOST and ASIAA (Taiwan), and KASI (Republic of Korea), in cooperation with the Republic of Chile. The Joint ALMA Observatory is operated by ESO, AUI/NRAO, and NAOJ.
The $F814W$-band image is 
based on observations made with the NASA/ESA Hubble Space Telescope, 
obtained from the data archive at the Space Telescope Science Institute. 
STScI is operated by the Association of Universities for Research in Astronomy, 
Inc. under NASA contract NAS 5-26555.
The National Radio Astronomy Observatory is a facility of the National Science Foundation operated under cooperative agreement by Associated Universities, Inc.

\appendix 

\section{X-CIGALE parameter}

Table A\ref{tab:cigale} lists the parameters adopted in {\sf X-CIGALE}.  The details of the parameters are described in \citet{2020MNRAS.491..740Y}.

\begin{deluxetable}{ll}
\tablenum{1}
\tablecaption{Range of X-CIGALE model parameters \label{tab:cigale}}
\tablewidth{0pt}
\tablehead{
\colhead{Component} & \colhead{Model}}
\decimalcolnumbers
\startdata
SFH        			& sfhdelayed model without no additional burst. $\tau_{\rm main} = $ 5, 10, 20, 50, 100, 200, 500, \& 1000 Myr, \\
					&  $\rm age_{\rm main} =$ 0.4 to 2.2 with 0.2 Gyr steps. \\
stellar population 	& \citet{2003MNRAS.344.1000B} SSP model, \citet{2003PASP..115..763C} IMF, and solar metallicity \\
dust attenuation 		& The modified \citet{2000ApJ...533..682C} attenuation model. $E(B-V)_{\rm lines} = 0$ to 1 with 0.2 steps. \\
                                     & $E(B-V)$ of stellar continuum is 0.44 times the $E(B-V)_{\rm lines}$. \\
dust emission             & \citet{2014ApJ...784...83D} dust emission model.  $\alpha=$ 0.0625, 0.2500, 1.0000, \& 2.0000. \\
				        & The AGN fraction of this component is set to zero. \\
nebular emission	& None \\
AGN (Fritz06)		& \citet{2006MNRAS.366..767F} model.  r\_ratio = 60.0, $\tau = 1.0$, $\Gamma = 0.0$, opening\_angle of the dust torus $= 100$,  angle \\
					& between equatorial axis and line of sight ($\psi$) = 0.001 to  60.100 with 10 deg steps (= 90 for type-1 and \\
					& 0 for type-2), AGN fraction = 0.0 to  1.0 with  0.1 steps, The extinction law of polar dust is \\
					& \citet{2000ApJ...533..682C},  E(B-V) for extinction in polar direction $=$  0.2, 0.4,~\& 0.6, temperature  of the polar\\
					& dust  $=  100$ K, and emissivity index of the polar dust $= 1.6$. \\
X-ray              		& AGN photon index $\Gamma = 0.4$ to 2.0 with 0.2 steps,  maximum deviation of alpha\_ox from the empirical  \\
					&  relation $= 0.2$, The photon indices of LMXB and HMXB are 1.56 and 2.0, respectively. \\
radio                            & Power law slope $\alpha=$ 0.4, 0.6, 0.8, \& 1.0, radio-IR correlation $q_{\rm IR} =$ 0.01, 0.1, 0.3, 0.5, 1.0, \& 2.5.  \\
free parameters		& $\tau_{\rm main}$, $\rm age_{\rm main}$, and $E(B-V)_{\rm lines}$ for stellar population. AGN fraction and $\psi$ for AGN (Fritz06) model. \\
                                     & $\Gamma$ for AGN (X-ray) model. $\alpha$ and $q_{\rm IR}$ for radio model. Thus free parameters are eight in total. \\
\enddata
\end{deluxetable}

\bibliographystyle{aasjournal}



\end{document}